\documentclass[preprint,a4paper, number,times,sort&compress,12pt]{elsarticle}
\journal{Computer Physics Communications}
\usepackage[utf8x]{inputenc}
\usepackage{graphics,graphicx}
\usepackage{epstopdf}
\usepackage{amsmath}
\usepackage{listings}
\usepackage{natbib}
\usepackage{xcolor}
\usepackage{hyperref}

\newcommand{\ai}{{\it ab initio}}

\definecolor{bluekeywords}{rgb}{0.13,0.13,1}
\definecolor{greencomments}{rgb}{0,0.5,0}
\definecolor{redstrings}{rgb}{0.9,0,0}
\definecolor{listingsbg}{HTML}{EFEFFF}

\begin{document}

\begin{frontmatter}

\title{A parallel algorithm for Hamiltonian matrix construction in electron-molecule collision calculations: MPI-SCATCI}
\author[ucl]{Ahmed F. Al-Refaie\corref{cor1}}
\ead{ahmed.al-refaie.12@ucl.ac.uk}

\author[ucl]{Jonathan Tennyson}
\ead{j.tennyson@ucl.ac.uk}

\address[ucl]{Department of Physics \& Astronomy, University College London, Gower Street, London WC1E~6BT, United Kingdom}

\begin{abstract}
  Construction and diagonalization of the Hamiltonian matrix is the
  rate-limiting step in most low-energy electron -- molecule collision
  calculations.  Tennyson (J Phys B, 29 (1996) 1817) implemented a
  novel algorithm for Hamiltonian construction which took advantage of
  the structure of the wavefunction in such calculations. This
  algorithm is re-engineered to make use of modern computer
  architectures and the use of appropriate diagonalizers is
  considered. Test calculations demonstrate that significant speed-ups
  can be gained using multiple CPUs. This opens the way to
  calculations which consider higher collision energies, larger
  molecules and / or more target states. The methodology, which is
  implemented as part of the UK molecular R-matrix codes (UKRMol and
  UKRMol+) can also be used for studies of bound molecular Rydberg
  states, photoionisation and positron-molecule collisions.
\end{abstract}

\end{frontmatter}

\section{Introduction}

Modelling low-energy electron-molecule scattering systems is vital to
the understanding of a range of physical processes in fields such as
plasma physics \cite{16BaKu}, astrophysics \cite{jtTF}, cell and DNA
damage \cite{bouda}.  There are a number of codes available for
performing \ai\ calculations on such collisions.  The most general of
these rely on use of the so-called close-coupling expansion where the
scattering wavefunction, $\Psi_k$, is represented, at least in the
region of the molecular target, by:
\begin{equation}
\label{eq:rmat_wf}
\Psi_k =  {\cal{A}}\sum_{ijn}\phi_{in}(x_1,x_2,....x_N)u_{ij}(x_{N+1})a_{injk} +
\sum_\ell\chi_{\ell}(x_1,x_2,...x_{N+1})b_{\ell k}
\end{equation}
where $\phi_{in}$ are the target wavefunctions and $u_{ij}$ are
continuum orbitals.  The index $i$ is the target symmetry, $j$ is the
continuum orbital index and $n$ counts over target states belonging to
symmetry $i$. $\cal{A}$ is the anti-symmetrization operator to ensure that
the target times continuum wavefunction obeys the Pauli principle.
The $\chi_{\ell}$ are short-range or $L^2$ functions where all electrons
occupy target orbitals; $x_p$ represents the coordinates of electron
$p$ where it is assumed that the target has $N$ electrons and hence
the scattering system has $N+1$ electrons. Finally, $a_{injk}$ and $b_{lk}$ are
the variational coefficients obtained by diagonalization of the
Hamiltonian.

A variety of different models can be represented by this
close-coupling expansion \cite{jt646}, including ones based on the use
pseudo-states to augment expansion in physical target states.
This approach is employed
in the R-matrix with pseudo-states (RMPS) procedure
\cite{B98,jt341,jt354}. In general, the target wave-function is
expanded in configuration interaction (CI) form as a linear
combination of configuration state functions (CSFs) $\eta$:
\begin{equation}
 \phi_{in} = \sum_m c_{imn}\eta_{im}.
\end{equation}

The Hamiltonian matrix derived from the use close-coupling expansion
described above has a characteristic structure, see
Fig.~\ref{fig:MatrixClasses} below.  In 1996, Tennyson \cite{jt180}
showed that it was possible to exploit this structure to greatly
speed-up the construction of scattering Hamiltonians. His algorithm,
as implemented in module SCATCI and which is discussed in detail
below, has formed the backbone of various implementations of the UK
Molecular R-matrix codes \cite{jt161,jt225,jt518,jt416,z.ukrmol+}.  The algorithm
used by SCATCI is extremely efficient leading it to be used for
extensive close-coupling calculations on electron collisions N$_2^+$
\cite{jt574}, SiN \cite{Kaur2015}, CH$_3$CN \cite{jt610},
uracil \cite{jt459,14mg}
pyramidine \cite{Mason2016} and many others systems, as well as for
studies of positron--molecule collisions \cite{jt411,jt491,jt510}.

Hamiltonian construction and diagonalization is usually the slowest
step in the \ai\ treatment of low-energy electron-molecule scattering.
While diagonalization can usually spread over a number of cores,
SCATCI is currently limited by its serial nature. This step in the
calculation can become expensive if one or more of the following
applies: (a) the use of an extensive list target states and/or
pseudo-state; (b) large target CI expansions; (c) large continuum
orbital basis sets.  With modern calculations it is quite possible
that all three of these criteria apply and it is quite easy to design
desirable but intractable calculations for even diatomic targets
\cite{jt444}. The recent development of the B-spline based UKRMol+
code \cite{z.ukrmol+,jt682}, which extends both the range of energies
and size of target that it is possible to treat, has only further
exacerbated this situation.

It would therefore appear timely to revisit Tennyson's Hamiltonian
build algorithm. This algorithm was designed to run on serial
computers and, given the prevalence of modern multi-core
architectures, it is on a parallel implementation that we particularly
focus. At the same time, it has been recognized that
the structure of various the Hamiltonian matrices that can be
generated using this algorithm lend themselves to different
diagonalization procedures \cite{jt332,jt510}. These different
diagonalization options are also integrate into a new MPI code,
MPI-SCATCI. The next section specifies the formal aspects of the new
code. Section 3 introduces the code itself, which is freely available
from the ukrmol-in project in the CCPForge program
depository\footnote{\url{https://ccpforge.cse.rl.ac.uk/gf/project/ukrmol-in}  (registration required)},
Section 4 gives some illustrative timings. Our conclusions and
suggestions for future work are given in Section 5.

\section{Theory}

Using the close-coupling expansion in the form of Eq.~(1) and
the target CI expansion of Eq.~(2) to build the final Hamiltonian
matrix one needs, at least in principle, 
to evaluate many Hamiltonian matrix elements of the form
\begin{equation}
\label{eq:uncontracted}
 H_{imj,i'm'j'} = \langle\eta_{im}u_{ij}|\hat{H}|\eta_{i'm'}u_{i'j'}\rangle
\end{equation}
In practice the CSFs are expanded in terms of an orthogonal set of
spin-orbitals which in turn are represented by atom and molecule
centred functions of various types \cite{jt474}, which are discussed
further below.  However, we note that for accurate calculations, the target
CI expansions can be very long and the the number of target states is
usually significantly smaller than the number of target CSFs. To
allow the identification of target states within the expansion, for
example to correctly assign asymptotic channels, it is
desirable to contract the matrix:
\begin{equation}
\label{eq:contracted}
 \tilde{H}_{inj,i'n'j'} = \sum_{m,m'} c_{imn}c_{i'm'n'}H_{imj,i'm'j'}
\end{equation}
where $c_{imn}$ is the coefficient of the target CI expansion, see Eq.~(2).
The target-$L^2$ part of the calculation undergoes a similar contraction
scheme:
\begin{equation}
 \tilde{H}_{inj,l} = \sum_{m} c_{imn}H_{imj,l}.
 \label{eq:contconL2}
\end{equation}
These steps very significantly reduce the size of the Hamiltonian,
hence the description contraction, and the majority
of matrix elements retained are usually between the $L^2$ functions.

\subsection{Symbolic evaluation, prototyping and expansion}

Whilst the matrix itself is smaller, evaluating Eq.
(\ref{eq:contracted}), in principle, requires the evaluation of all
integrals in Eq.  (\ref{eq:uncontracted}). SCATCI implements the
algorithm of Tennyson \cite{jt180} that avoids evaluating integrals
and explicit computation of the un-contracted Hamiltonian. This is
done through the manipulation of symbolic matrix elements \cite{ly81}
generated using prototype CSFs \cite{Y73,jt140}. This procedure
exploits the structure of the scattering wavefunction, see Eq.~(1),
where a particular target wavefunction or CSF is multiplied by a (long)
list of continuum functions.

The Hamiltonian construction is driven from a list of CSFs.  The
required integrals are identified by the application of Slater's rules
to give a list of symbolic matrix element. The published versions of
the UKRMol/UKRMol+ codes are based on a traditional algorithm
for Slater's rules. Scemama and Giner \cite{13ScGi} proposed an
alternative algorithm that represents all possible spin orbitals that
an electron can occupy as an array of 64-bit integers. Each integer
can represent 64 spin orbitals and each set bit in the $n^{\rm th}$ position
represents an electron occupying the $n^{\rm th}$ spin orbital. Determining the
substitutions between two determinants can be easily computed by
performing an exclusive-or (XOR) followed by a population count
(popcnt). Their algorithm also determines the resultant phase from
possible spin orbital reordering by computing the number of occupied orbitals between the differing orbitals using bit masks.
A bug fix was applied to their optimized implementation that caused incorrect masks to be generated when orbital indices where of a multiple of 64, this fix has since been applied to the original authors source code.
The majority of the computation relies mainly on hardware native instructions such as bit shifts, XOR
and popcnt (AVX+ instruction set) making it extremely efficient and
reducing overall Hamiltonian build times by factors of two to five.
This algorithm is also independent of the number of electrons making
it ideal for large polyatomic molecules. This efficient algorithm is the one
used in MPI-SCATCI

To briefly summarize, the SCATCI
procedure involves transforming the evaluation of both
Hamiltonian into symbolic form. For Eq. (\ref{eq:uncontracted}) this is
of the form:
\begin{equation}
\label{eq:symbol:uncontracted}
 H_{imj,i'm'j'} = \sum_\alpha C^{\alpha}_{imj,i'm'j'}X(I^{\alpha}_{imj,i'm'j'})
\end{equation}
Where $\alpha$ are the associated integral indices $I^{\alpha}$ and
coefficients $C^{\alpha}$ generated for the matrix element and $X$ is
the integral function.  Within the UK molecule R-matrix codes,
the  indices are four 16-bit integers representing
the associated orbitals in the integral packed into a 64-bit integer.
As the configurations obey the Slater-Condon rules, there are a fixed
number of one and two electron integrals that can be precomputed ahead
of time. Therefore, $X$ is
a function that maps the indices $I$ into a one and two-electron
integral array and returns the appropriate integral value.

The contracted matrix can also be expressed similarly:
\begin{equation}
\label{eq:symbol:contracted}
 \tilde{H}_{inj,i'n'j'} = \sum_\beta
D^{\beta}_{inj,i'n'j'}X(I^{\beta}_{ij,i'j'})
\end{equation}

Whilst there are essentially no integral evaluations, large numbers of
continuum orbitals and target configurations make this computationally
undesirable. One way this is circumvented is to utilize symbolic
prototyping.  This methodology removes the need to
explicitly evaluate matrix elements for each continuum orbital $j$ by
instead evaluating the full Hamiltonian matrix elements for one or two
prototype configurations corresponding to one or two $j$ and
generating the full symbolic lists by manipulating the integral
indices. This is the essence of Tennyson's algorithm \cite{jt180}.

Transformation the full Hamiltonian matrix to the CI contracted one can be
performed by contracting the minimal prototype symbolic elements of
$H$. The prototype integral labels do not change but the associated
coefficients do depending on the target symmetries and states:
\begin{equation}
 D^{\beta}_{inj,i'n'j'} = \sum_{m,m'}c_{imn}c_{i'm'n'} C^{\alpha}_{imj,i'm'j'}
\label{eq:contract-sym}
\end{equation}
after which the labels are expanded into the full range of $j$. This means that
the full Hamiltonian is never explicitly evaluated.

\subsection{Matrix classes}

\begin{figure}
\centering
\includegraphics[width=1.0\textwidth]{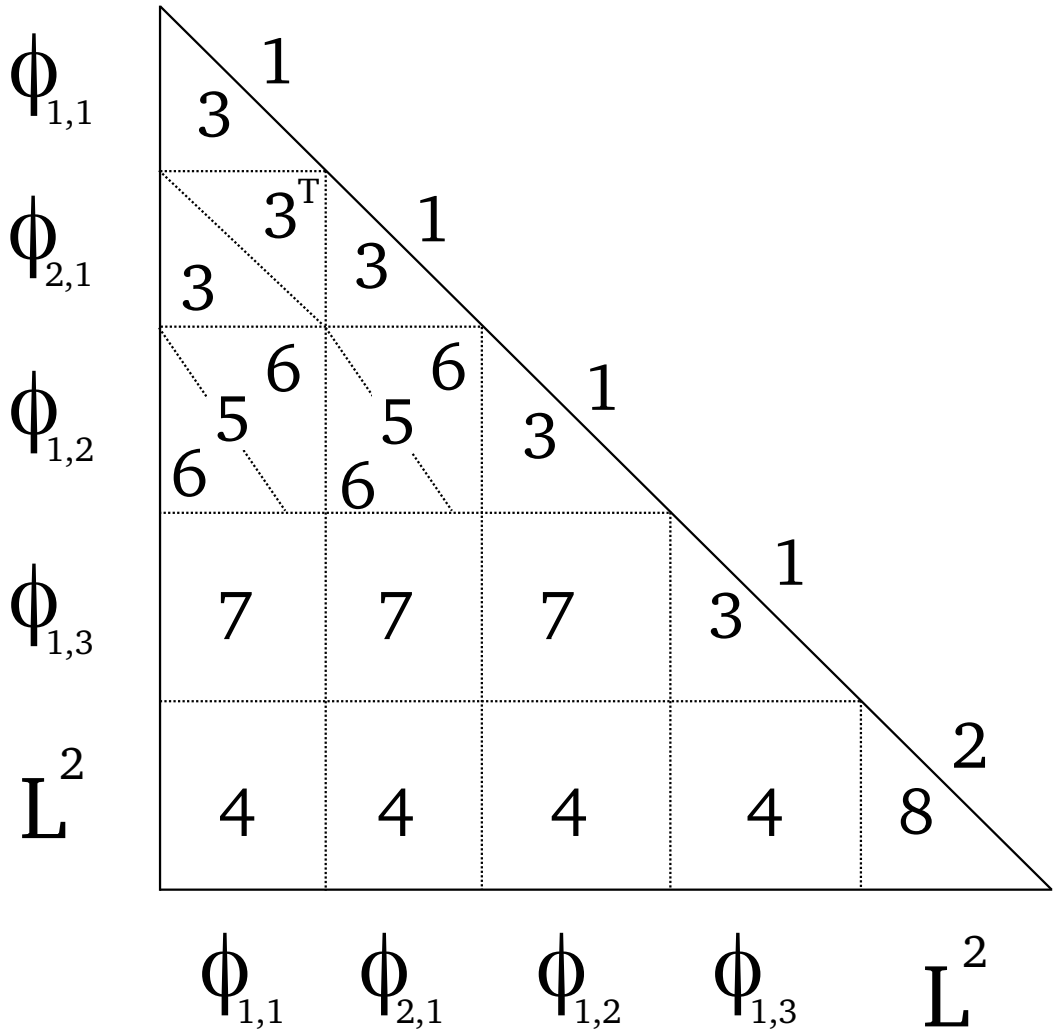}
\caption{Class structure of the contracted Hamiltonian for a scattering
calculation with three target symmetries. $\phi_{i,j}$ denotes a target state $i$ for a target symmetry $j$. Target symmetries 1 and 2 couple to
the same continuum symmetry and symmetry 1 has two target states $\phi_{1,1}$
and $\phi_{2,1}$.}
\label{fig:MatrixClasses}
\end{figure}

The consequence of this contraction is that the matrix is now split into
differing contraction classes based on the symmetry properties of the target
states and the $L^2$ functions. Below is a summary of these contracted classes
with Figure \ref{fig:MatrixClasses} illustrating them. Since the Hamiltonian is
real symmetric, only the lower triangular portion is considered.

\subsubsection{Classes 1 and 3}
Classes 1 and 3 involve matrix elements between functions with the
same target symmetry.  Class 1 are the diagonal matrix elements involving a target state times a continuum orbital.
These integrals can also occur in off-diagonal elements involving different
states of the same symmetry. Class 3 are the off-diagonal elements involving
different target states of a given symmetry; they have a symmetric block structure. The upper triangular block is the
transpose of the lower triangular. The contraction is of the form:
\begin{equation}
 \tilde{H}_{inj,in'j'} = \sum_{m,m'}c_{imn}c_{imn'}H_{imj,imj'} + 
\sum_{m,m'}(c_{imn} + c_{im'n'} + c_{im'n}c_{imn'})H_{imj,im'j'}
\end{equation}
The symbols are then expanded for all target states and continuum orbitals of
the target symmetry.
\subsubsection{Classes 2 and 8}
Classes 2 and 8 are the diagonal and off-diagonal matrix elements of the $L^2$ functions. These
undergo no form of contraction and are instead evaluated explicitly. These are
referred below to as `pure $L^2$' elements.
\subsubsection{Classes 5 and 6}
These classes involve elements where the target symmetry differs but their
coupled continuum symmetry is the same. Class 5 is the diagonal element of the
local matrix block and class 6 is the off-diagonal element. The contraction
scheme is of the form:
\begin{equation}
 \tilde{H}_{inj,i'n'j'} = \sum_{m,m'}c_{imn}c_{i'm'n'}H_{imj,i'm'j'} 
\label{eq:contract56}
\end{equation}

\subsubsection{Class 7}
This class involves matrix elements where both the target and
continuum symmetry differs. The contraction is the same as Eq.
(\ref{eq:contract56}) and is expanded across both target states and
continuum orbitals of the target symmetries
\subsubsection{Class 4}
For matrix elements between continuum and $L^2$ functions the contraction is
one-dimensional and is of the form given by Eq. (\ref{eq:contconL2}).

\subsubsection{Sparsity and diagonalization}

\begin{figure}
\centering
\includegraphics[width=1.0\textwidth]{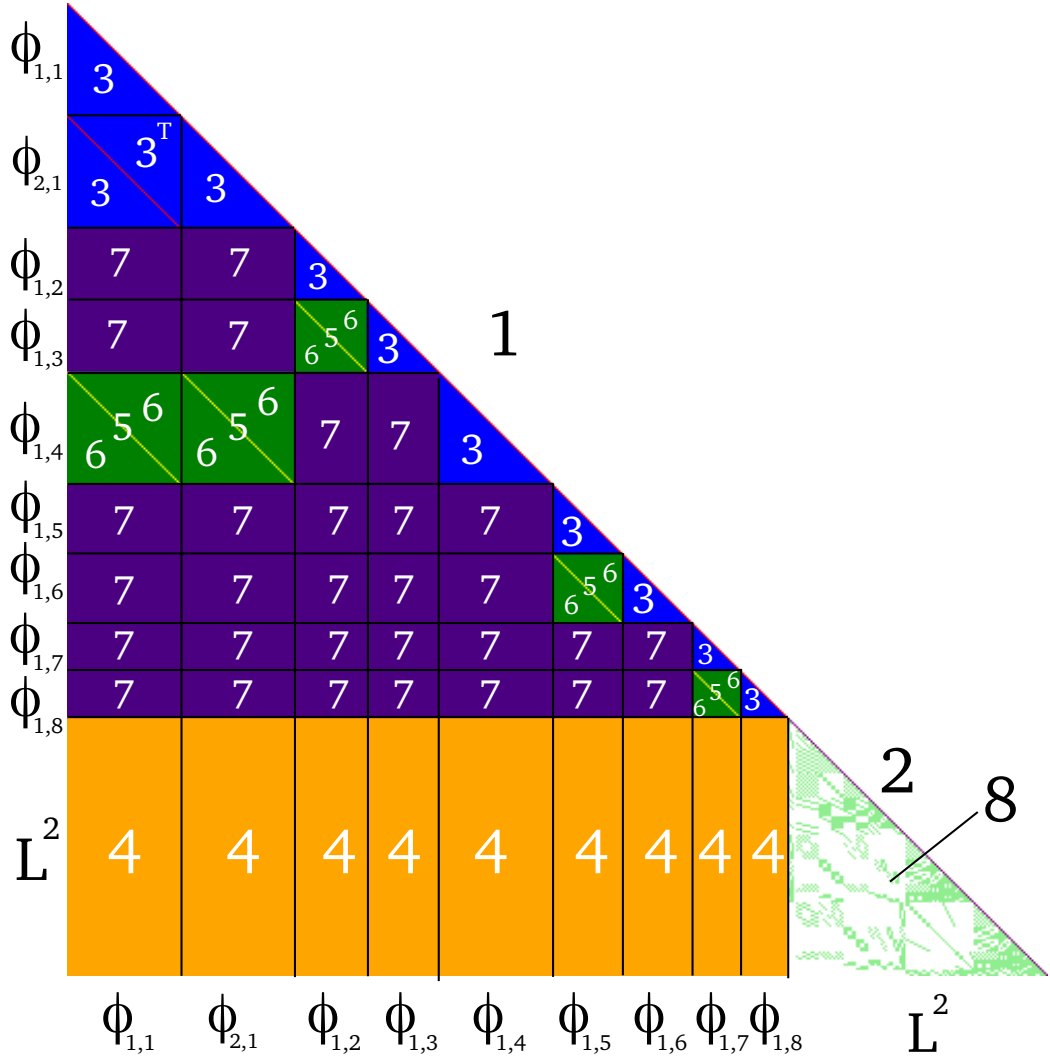}
\caption{The lower-triangular Hamiltonian output for a test case for H$_2$O with
8 target symmetries and 4 continuum symmetries. Symmetry 1 has two target
states. Target symmetries 1 and 4 both couple to the continuum symmetry 4,
denoted by $(1,4)\rightarrow 4$, leading to the appearance of contraction classes 5 and 6 in the matrix blocks ($\phi_{1,4}$,$\phi_{1,1}$)
         and ($\phi_{1,4},\phi_{2,1}$). Other couplings include: ~$(2,3)\rightarrow 3$,~$(5,6)\rightarrow 2$,~$(7,8)\rightarrow 1$. White
colors represent zero values. }
\label{fig:watertest}
\end{figure}

For almost all classes, the matrix blocks are dense with the exception
of the off-diagonal pure-$L^2$ class which is sparse. An example test
case calculation of electron-H$_2$O \cite{jt281} given in Figure
\ref{fig:watertest} illustrates these properties. The nature of the
matrix changes depending on the choice of CSF. For this test case, the
matrix is considered dense due to the smaller number of $L^2$ functions,
which encourages the usage of a dense diagonalizer such as
LAPACK\cite{99AnBaBi.method}. When dealing with large scattering
systems of interest such as phosphoric acid, the number of $L^2$
functions can be several orders of magnitude larger than the size of
the contracted part of the matrix. The matrix at this point becomes
extremely sparse and may necessitate usage of a sparse diagonalizer
such as ARPACK\cite{arpack.method}.

\subsection{The UK-molecular R-matrix codes}
The UK-molecular
R-matrix suite, UKRMol \cite{jt225}, and its new enhanced
version, UKRMol+ \cite{z.ukrmol+}, are a set of modules and programs to fully
solve the electron/positron-molecule scattering problem using the R-matrix
methodology. The suite splits into two sets of codes:
UKRMol-in\footnote{\url{https://ccpforge.cse.rl.ac.uk/gf/project/ukrmol-in}}
which deals with the inner-region problem and
UKRMol-out\footnote{\url{https://ccpforge.cse.rl.ac.uk/gf/project/ukrmol-out}}
which deals with the outer region. 

The codes use a a variety of methods to represent the target and
continuum functions. The original code  \cite{jt161} only consider
electron collisions with diatomic molecules; it was based on the
use of Slater type orbitals (STOs) to represent the target wavefunction
and suitably orthogonalised \cite{jt61} numerical functions for the
continuum. The original polyatomic code \cite{jt204,jt225} uses
Gaussian type orbitals (GTOs) for both the target and the 
continuum functions. The new UKRMol+ \cite{z.ukrmol+} also uses GTOs to
represent the target but uses a hybrid GTO -- B-spline basis set for the
continuum. One effect of this is that the continuum expansions can become
significantly larger as the code allows treatment of more partial waves
(higher $\ell$ values), larger R-matrix spheres and extensions to higher
energies, all of which lead to an increase in the number of continuum
functions.

A crucial module in both UKRMol and UKRMol+ is SCATCI \cite{jt180}.
SCATCI is a Fortran 77 code that deals with the building and
diagonalization of the inner-region $N+1$ scattering Hamiltonian and
is the last step before moving into the outer-region portion of the
calculation.

\section{MPI-SCATCI}

MPI-SCATCI is a complete rewrite of SCATCI in Modern Fortran (2003)
that uses MPI to perform both the N+1 Hamiltonian build and
diagonalization. Its design is heavily based on an Object-Orientated Programming 
(OOP) paradigm to give the code a high degree of flexibility for further modification. 
Whilst previously integrating
features such as a new integral format required a fair amount of
modification to build subroutines, the OOP approach allows simply for
the definition of a new Integral class with appropriate procedures
that is then attached to the Hamiltonian at run time without touching
the build code. Similar functionality also applies to the
diagonalizers and as will be discussed later on in Section
\ref{sec:Matrix}, gives MPI-SCATCI the ability to support almost any
diagonalizer library and matrix format with little to no modification.

\subsection{Build parallelization}

There are three avenues for parallelisation of the scattering Hamiltonain:
Across prototyping and contraction, across the expansion and across the L$^2$
functions. The choice is dependant on the type of matrix class being calculated.
Regardless of which method is used, every process performs the same class and
the same target symmetry for the contracted classes as it simplifies the
distribution of work.

\subsubsection{Classes 2 and 8}

\begin{figure}
\centering
\includegraphics[width=0.5\textwidth]{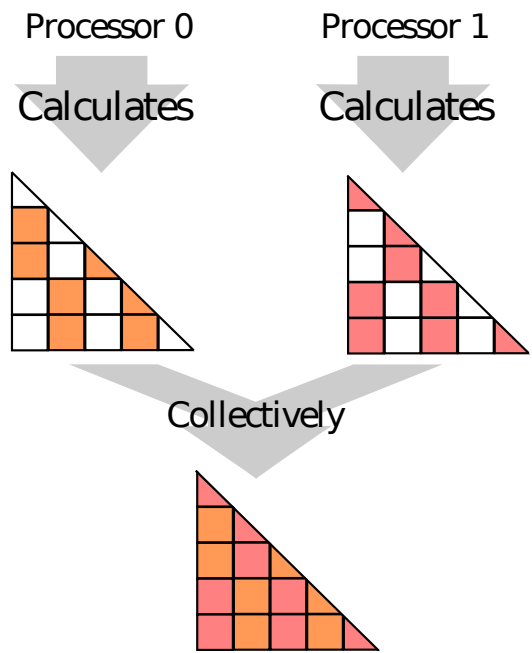}
\caption{A visual representation of two processes evaluating the matrix elements
of classes 2 and 8 in the scattering Hamiltonian}
\label{fig:l2calc}
\end{figure}

These pure-$L^2$ elements are the most straightforward to consider as no
contraction is required. Each Slater-Condon calculation gives symbols
for a single matrix element. MPI-SCATCI combines both classes into a
single calculation and the nested loop over the lower triangular is
collapsed into a single loop. This loop is evenly distributed across
all MPI processes and computed independently with no need for any form
of communication. Viewing the calculation as a whole (as seen in
Figure \ref{fig:l2calc}), 'collectively' all matrix elements have been
calculated. This is the parallelisation that is also used in the construction
of the target Hamiltonian whose solutions are required to give the
target wavefunction coefficients, see Eq.~(2), and associated energies,
see below.

\subsubsection{Classes 1, 3, 5, 6, 7 }

\begin{figure}
\centering
\includegraphics[width=1.0\textwidth]{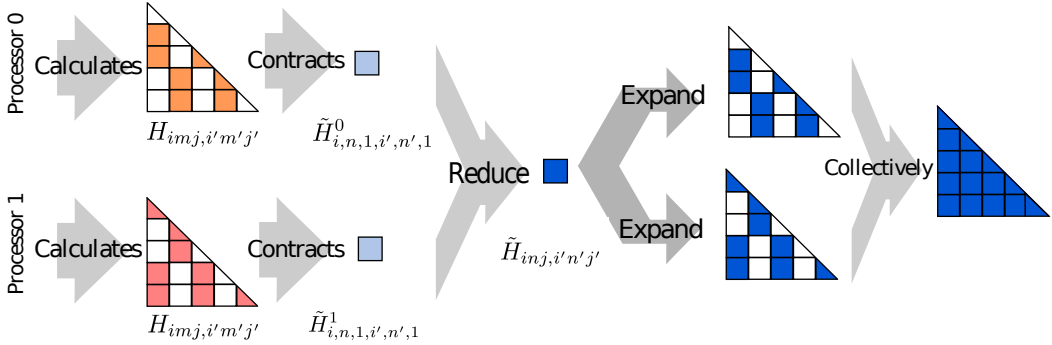}
\caption{A visual representation of two processes evaluating the matrix elements
of continuum only classes. The first stage involves computing and contracting
the prototype elements into the first incomplete symbolic elements of the
contracted matrix. The second stage is the all process reduction into the full
symbolic prototype list. The last stage is the expansion across $j,j'$ for each
process. This means that both processes collectively have the full contracted
matrix. }
\label{fig:class13567}
\end{figure}

For pure continuum classes, there is a two step parallelization that occurs.
Figure \ref{fig:class13567} visually describes this process. The first step
consists of parallelizing the prototype and contraction stage. The loop across
prototype CSFs are evenly distributed among MPI processes. This means each
processor $p$ produces an incomplete list of contracted prototype symbols
$\tilde{H}^{p}_{inj,i'n'j'}$.
At this point, a gather and reduce is performed on all $P$ processes in order to
retrieve the complete set of contracted prototype symbols:
\begin{equation}
 \tilde{H}_{inj,i'n'j'} = \sum_p^{P} \tilde{H}^{p}_{inj,i'n'j'}
\end{equation}
Whilst this is a significant synchronization step, it is only performed 
\begin{equation}
 \frac{n_{t}(n_{t}+1)}{2} 
\end{equation}
times where $n_t$ is the number of target symmetries. This, in general, is 
small number of synchronization points as the number of
target symmetries in a calculation is rarely larger than single digits.

The second step is the expansion of the prototype symbols. Very little
work is actually done during the expansion as it only requires
modifying the indices of the integral label for each continuum
orbital. However, some systems have target symmetries that are coupled to
thousands of continuum orbitals. Their prototypes require expanding for millions of
matrix elements for the off-diagonal classes 3, 5, 6 and 7. Originally each
process performed the full expansion on their incomplete symbols
followed by a reduction to retrieve the full matrix element. This
unfortunately became a significant bottleneck for continuum heavy
systems and therefore necessitated the need for parallelizing the
expansion process as well. The parallelization of the expansion is
performed by evenly distributing processes across the $j,j'$ expansion
loop. Essentially, this step behaves similar to classes 2 and 8 and
collectively the full matrix elements are computed.  This two step
approach benefits two types of problem sizes. For systems with a large
number of prototype CSFs, the first step gives the greatest gain in
performance. For systems with target symmetries that contain a huge
number of continuum orbitals, the second stage provides the greatest
benefit.

The original SCATCI had the option to remove all integrals involving
only target orbitals from the matrix elements of classes 1, 3, 5, 6 and 7. 
These are replaced by adding
the appropriate (precomputed) target
energy along the diagonal. This approach has a number of advantages: it
significantly reduces the number of integral evaluations and facilitates
the manipulation of target energies, see Ref.~\cite{jt208} for example.
In future we plan to use this facility, which is retained
in MPI-SCATCI, to simplify the treatment heavy atoms via the use of
effective core potentials.

As discussed by Orel {\it et al.} \cite{orl91}, there can be a technical
issue with phases of the wavefunctions generated in a pure target
calculation (Eq.~(2)) and the scattering wavefunction (Eq.~(1)) to do with
order in which the electrons are treated in the CSFs. Ignoring this
problem has in the past led to generation of incorrect results,
see Gillan  {\it et al.} \cite{jt179}.
SCATCI resolves this problem by using a phase mask which matches
the phases between the two wavefunctions \cite{jt195}. This is
retained in MPI-SCATCI.

\subsubsection{Class 4}

Class 4 is the off-diagonal continuum-$L^2$ portion of the matrix.
This presents a problem as it requires prototyping and expansion for
each target symmetry and $L^2$ function. The number of $L^2$ functions
in a typical calculation is significantly larger than size of the
contracted continuum. This presents a problem with the two step
procedure as each $L^2$ function for each symmetry would require a
prototype synchronization step resulting in possibly millions in order
to complete.

\begin{figure}
\centering
\includegraphics[width=1.0\textwidth]{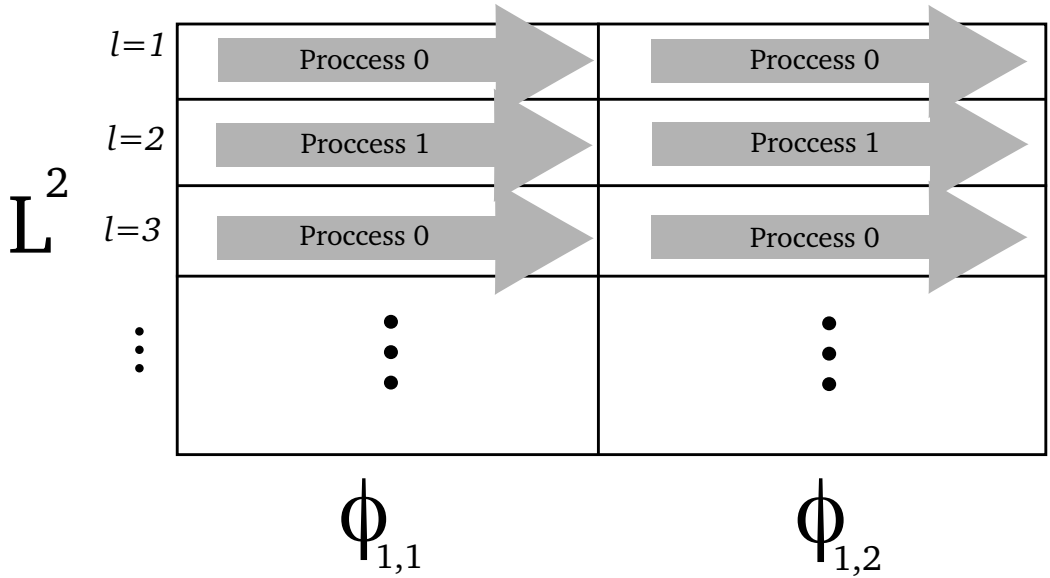}
\caption{A visual representation of the class 4 prototyping and expansion of a
single target symmetry block containing two states using two processes. The
arrows represent the direction of generation of matrix elements. The processes
are distributed in a round robin style across each $\ell$ function then prototype,
contract and expand for all target states in the symmetry.  }
\label{fig:class4}
\end{figure}

However we can exploit the fact that the prototyping, contraction and
expansion are all one-dimensional loops as given by Eq.
(\ref{eq:contconL2}). This means that a class calculation for each
$L^2$ function is significantly easier and faster to perform than any
other contracted class. With this, the parallelization is instead
performed across $\ell$ functions for a specific target symmetry
independent of any other process and eliminating any costly
synchronization. This is illustrated in Figure \ref{fig:class4}

\subsection{Matrix distribution}
\label{sec:Matrix}
One of the key goals for the MPI code is to not only support the diagonalizer
already included in the serial SCATCI code but also a wide range of MPI
diagonalizers. Since often all eigenvectors are required, a dense diagonalizer
such as SCALAPACK \cite{slug} is ideal for its efficient householder approach.
However, for scattering Hamiltonians that have sizes in the order of millions,
an iterative diagonalizer, such as the Scalable Library for Eigenvalue Problem
Computations (SLEPc)\cite{Hernandez:2005:SSF}, is more desirable as matrix
sparsity can be exploited for storage and often these sizes arise in partitioned
R-matrix problems that require only a small percentage of eigenvectors.

The Hamiltonian should therefore be built with regards to the final
processor arrangement of the matrix elements. However both SCALAPACK
and SLEPc have vastly different methods of storing the matrix.
SCALAPACK uses a block-cyclic distribution and SLEPc uses a blocked
row distribution. In addition, SLEPc requires the upper triangular
matrix in C-style indexing. Therefore separate Hamiltonian builds
would be required for each type of matrix distribution. This presents
a huge cost in time to write, test and debug every Hamiltonian for
every distribution.

\subsubsection{Matrix formats}

In MPI-SCATCI, an abstract \verb+BaseMatrix+ class is defined which
provides the Hamiltonian a standard routine to store a matrix element.
An inherited abstract \verb+DistributedMatrix+ class is also defined
which will distribute the matrix elements in any format. It works by
allowing the processes to touch every single matrix element at least
once at some point during execution and by applying rules.

\begin{figure}
\centering
\includegraphics[width=1.0\textwidth]{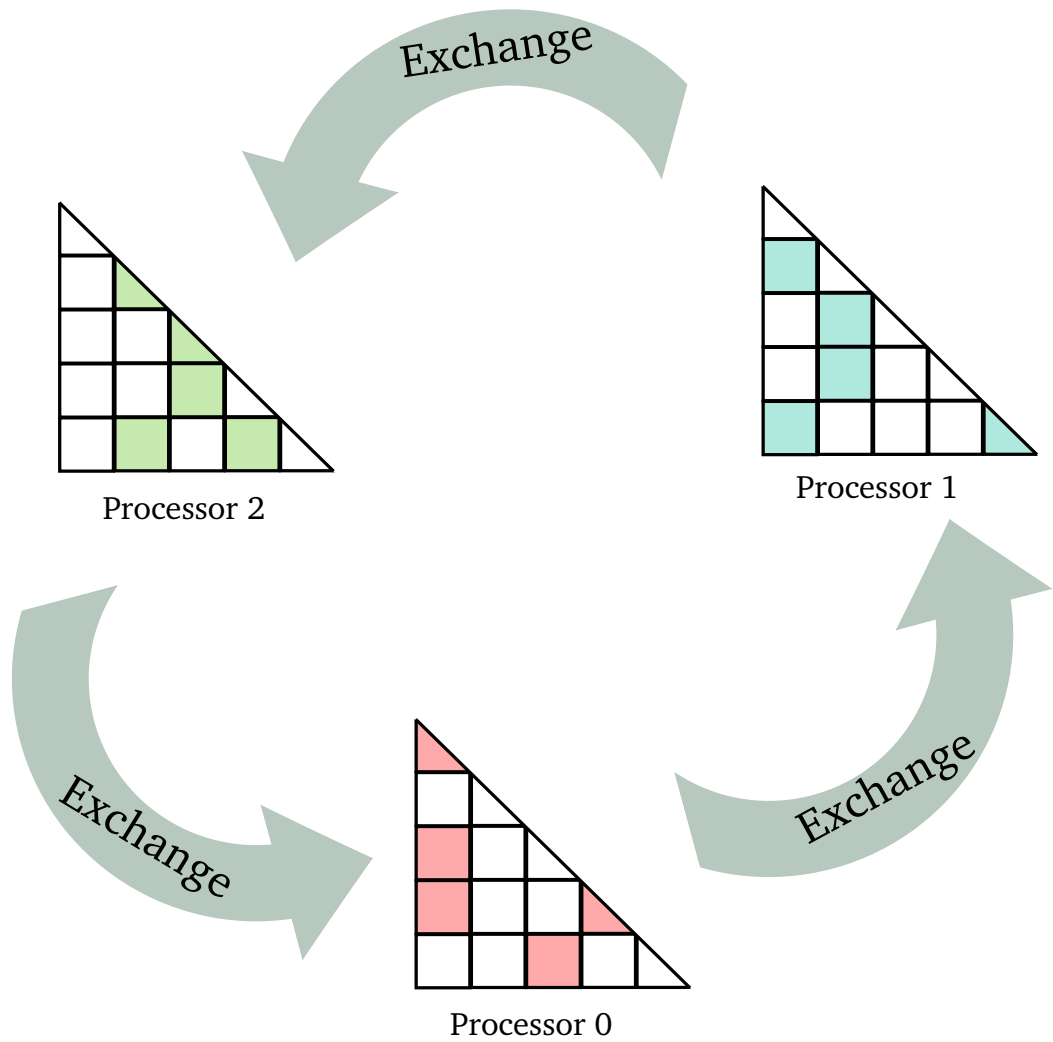}
\caption{Ring rotation of the each processes hot storage}
\label{fig:L2-update}
\end{figure}

The rules are defined by a virtual boolean function. Each inherited matrix
format must define this function in order to properly place matrix elements into
the correct process.
Two types of storage are defined, hot and cold storage. Hot storage is temporary
storage for matrix elements that, after applying rules, do not belong to the
computing process. This type of storage is the same for all distributed matrix
class. Cold storage is the permanent storage of matrix elements that will
eventually be used in diagonalization. Its representation depends on the format,
for SCALAPACK it is the local matrix array, for SLEPc it is a Petsc \verb+Mat+
object.
At some point during the build an update can be triggered. This update consists
of rotating the hot storage in a ring like fashion across every process as
illustrated in Figure \ref{fig:L2-update} and applying the defined rules to each
element for possible placement into cold storage. This means that communication
through an interconnect only occurs with 2 of the processors in a node giving a
communication overhead of the order $2(n-1)$ where $n$ is the number of nodes.
Once each processes hot storage has completed a circuit, it is cleared and ready
for usage again.

An update is triggered through two means: When memory has been
exhausted and when the Hamiltonian build is completed. MPI-SCATCI
tracks available free memory and when a process has exhausted all of
its available memory, this signals all other processes to begin an
update. This has the benefit in that when either given more RAM or
more processes, the number of updates required in a run decreases.
This is because we can either store more matrix elements or that we
are storing less matrix elements per process. As will be discussed in
Section \ref{sec:Perf} the update time remains essentially constant
across processor counts.

This \verb+DistributedMatrix+ has proved beneficial as, barring initialization,
requires only a single function definition to support the appropriate format
without touching the Hamiltonian build. Currently, MPI-SCATCI has defined three
types of distributed matrices:
\begin{itemize}
\item \underline{SCALAPACKMatrix}
  \begin{itemize}
  \item Stores local matrix in block cyclic distribution
  \item Used by SCALAPACK diagonalizer
  \item Distribution rule:
    \begin{itemize}
    \item Call INFOG2L. If it belongs to me then store and return true otherwise
false
    \end{itemize}
\end{itemize}
\item \underline{SLEPcMatrix}
  \begin{itemize}
  \item Stores in a PETSc Mat object in blocked row distribution
  \item Used by SLEPc diagonalizer
  \item Distribution rule:
    \begin{itemize}
    \item Convert into upper triangular matrix and C index format
    \item If it is within block row, store and return true otherwise false
    \end{itemize}
\end{itemize}

\item \underline{WriterMatrix}
  \begin{itemize}
  \item Writes matrix elements to file
  \item Used by original SCATCI diagonalizers
  \item Distribution rule:
    \begin{itemize}
    \item If I am master process, store (and write) and return true, otherwise
false
    \end{itemize}
\end{itemize}
\end{itemize}

The SLEPcMatrix class presents another interesting feature, the rule function
can also be used to preprocess a matrix element and in this case, convert into
upper triangular and C-style indexing before storing. Additionally, PETSc matrix
assembly time is non-existent as the elements are all in the correct process.

\subsubsection{Diagonalization}

The abstract \verb+Diagonalizer+ class provides a standard
diagonalization routine that accepts a \verb+BaseMatrix+ as input.
There is, however, no standard matrix element retrieval routine due to
the vastly differing ways the matrix is stored (and sometimes not
stored). Whilst it is technically valid to pass any \verb+BaseMatrix+
into any diagonalizer routine, it is up to the implementer to
determine how to access the element for each format. The currently
implemented diagonalizers perform a type check on the matrix pointer
and either halts if it is not supported or calls the correct
subroutine that typecasts back into its derived class. Whilst this
seems to go against the OOP approach used in the rest of the code, the
performance benefits are massive. For the SCALAPACK diagonalizer, the
SCALAPACKMatrix provides the correct local array to immediately begin
diagonalization and the same goes for the SLEPc diagonalizer and its
corresponding matrix.

It is also worth noting that both SLEPc and SCALAPACK can be used in
the same run. It is common for a run to use SLEPc to retrieve the
target coefficients for the Hamiltonian build to then shift to
SCALAPACK for the diagonalization of the scattering Hamiltonian.  The
diagonalizers currently supported by MPI-SCATCI are: LAPACK, Davidson
\cite{94sf}, ARPACK, SCALAPACK and SLEPc and all can be mixed in the
same run. Additionally, there is an experimental feature for non-MPI
diagonalizers to utilize all threads in a node by sleeping other MPI
processes whilst the master process performs OpenMP diagonalization.
This is beneficial for the parallel MKL LAPACK as it is generally more
efficient in a single node than SCALAPACK. However, this feature is
unreliable as it is dependant on the non-polling barrier
implementation of the MPI library and the pinning modes used.

\subsection{Large Integrals}

Under MPI, each process is given a private memory space. Taking an
example 24-core node with 64 GB of memory and distributing evenly,
this allows for a maximum of 2.5 GB per MPI process. Each process must
store its own local copy of the CSFs, the local matrix and the
integrals with this small amount of memory.  The biggest cost comes
from the integrals themselves. For instance, a the UKRMol phosphoric acid
scattering calculation considered below requires 1.5 GB of memory to
store the integrals leaving only 1 GB for everything else. This is a
bigger issue with UKRMol+ calculations using B-splines.  The integrals
for the recent electron-beryllium mono-hydride (BeH) UKRMol+
calculation \cite{jt682} require 3.0 GB, preventing them from being
used as one of our example systems. Scattering calculations on larger
systems such as uracil using B-splines may require tens of gigabytes
of memory to store all of the integrals.

\begin{figure}
\centering
\includegraphics[width=1.0\textwidth]{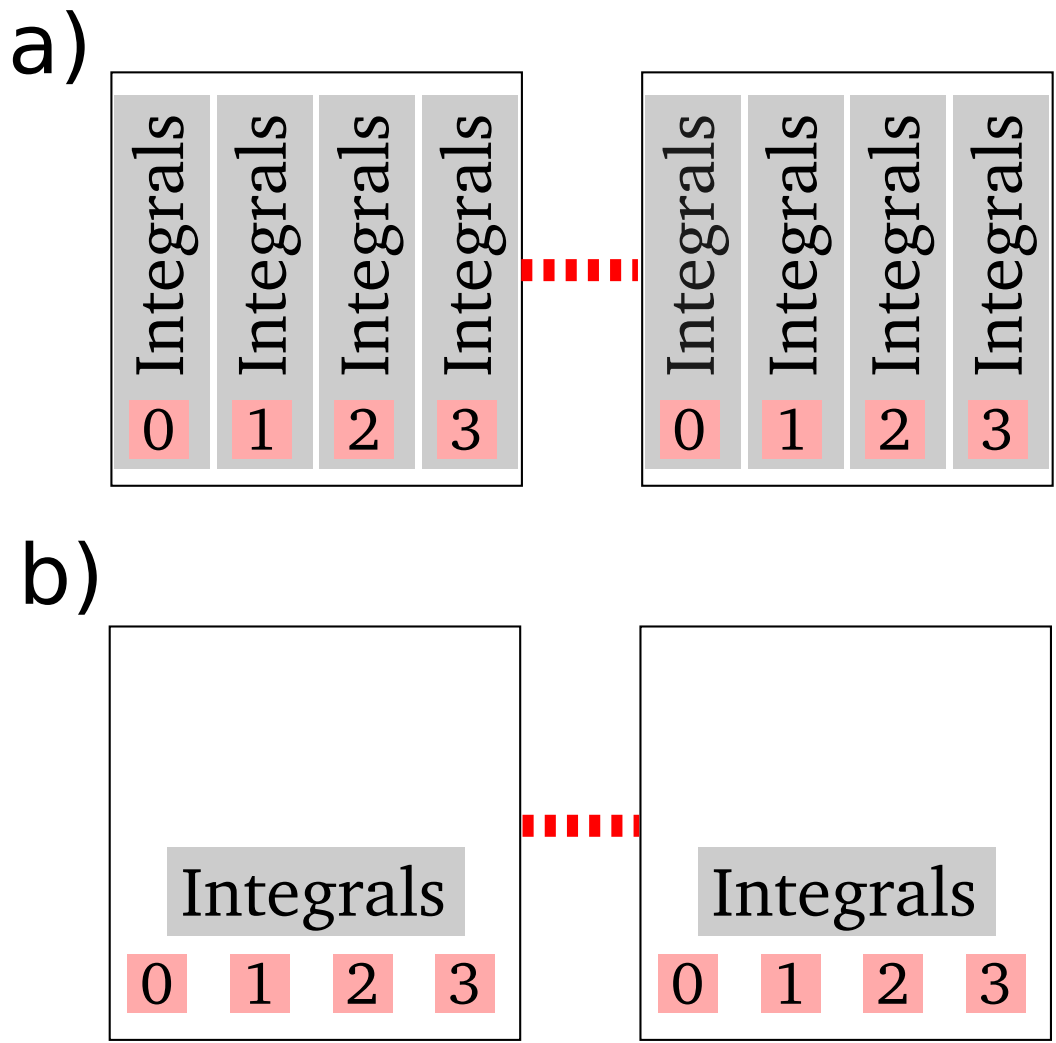}
\caption{Two Memory layouts of the integrals for two four-core nodes with interconnect
(red) between the two. a) The standard private memory layout under MPI and highlights the amount of repeated data on
each compute node. b) MPI-3.0 Shared memory layout, Only a single instance of the integral is
loaded for each node.}
\label{fig:integral-localmemory}
\end{figure}

The fundamental issue is that the integral data is being repeated
multiple times in each node as illustrated in Figure
\ref{fig:integral-localmemory}a. Naturally a method of distributing the
integrals is needed and there are many to choose from.

Firstly, an integral scatter method could be implemented where each
process has a portion of the integrals and are then moved across when
necessary. However it is difficult to predict which integrals are
needed by which process and this is compounded by the fact that
expansion of the prototype elements can introduce hundreds of
differing integrals that are not present within the computing process.
In a sense, at each integral evaluation it is likely that there will
be a significant degree of communication which will kill performance,
especially at higher process counts.

Another method would be to reduce the number of MPI processes in each
node but to then restore parallel performance by utilizing OpenMP. A
single occurrence of the integral could occur at each node giving us a
large amount of memory to store integrals as well as minimizing the
communication cost in synchronization as it will only occur at each
node rather than each process. However, OpenMP 4.0 support of modern
Fortran is still not mature enough and language features such as
polymorphism are not fully supported resulting in crashes.
Additionally some MPI diagonalizers are not hybrid OpenMP+MPI and therefore suffer in performance.

The method used by MPI-SCATCI utilizes a feature of the MPI-3.0 standard: Shared
memory

\subsubsection{Shared Memory}

Shared memory is a feature which allows a portion of memory to be seen
by a group of processes. This feature was actually present in some
form with the MPI-2.0 standard through the use of windows but was in a
sense a collective operation requiring many synchronizing 'epoch' in
order to get or put.  Additionally it is not aware of the
node-locality of certain processes and assumes remote memory access
(RMA) at all times. The MPI-3.0 standard introduces a new
communicator: \verb+MPI_COMM_TYPE_SHARED+. this communicator groups
process by which node they occur in. A shared window can be allocated
that is node aware and removes the need for RMA, improving memory
access times. Additionally there is no need for synchronization for
any puts or gets but is still necessary to ensure no race conditions
occur. However this is ideal as the integrals, once loaded, become
read only eliminating any further fencing on gets. The elimination of 
fencing also prevents costly cache synchronization steps though 
there are still local cache misses due to the random access nature of the integrals. 
The simplicity of this usage is a significant advantage as the shared memory array behaves
identically to a normal Fortran array once it has been set up. The
memory layout of our integrals now reflects the illustration in Figure
\ref{fig:integral-localmemory}b and affords the code the ability to
handle extremely large integrals that fit within the nodes total
memory.

% \begin{figure}
% \centering
% \includegraphics[width=1.0\textwidth]{integralmemoryshared.eps}
% \caption{Shared Memory layout of the integrals for two four-core nodes with
% fabric (red) connecting the two. Only a single instance of the integral is
% loaded for each node.}
% \label{fig:integral-sharedmemory}
% \end{figure}

A new module was created in order to facilitate this functionality. It replaces
the standard Fortran allocate function for arrays that we wish to share. For
example, allocating an array for the one electron integrals:
\begin{lstlisting}
allocate(one_e_int(num_one_e_int),stat=ifail)
\end{lstlisting}
Becomes this:
\begin{lstlisting}
one_electron_window = mpi_memory_allocate_real(one_e_int,num_one_e_int)     
\end{lstlisting}
If there is no available MPI-3.0 library, this will fallback into the standard
Fortran allocate routine. The window variable is used both for determining if
shared memory is being used and for deallocation. No other change in the
integral routines is necessary. No change in performance has been observed
between having a local private copy of the integrals and utilizing shared
memory.

\section{Performance}
\label{sec:Perf}

MPI-SCATCI has been successfully run and benchmarked on both University College
London's Grace@UCL supercomputing cluster and ARCHER, the UK National
Supercomputer Service. Grace@UCL has 360 nodes each with 16 Intel Haswell
cores and 64 GB of memory connected by non-blocking Intel Truescale Infiniband. ARCHER's Cray
XC30 nodes comprise of two 2.7 GHz, 12-core E5-2697
v2 CPUs with 32 GB each arranged in a non-uniform memory access (NUMA)
configuration giving 24 cores and 64 GB total connected with Cray Aries interconnect. 
The benchmark runs all stored the scattering Hamiltonian on disk rather than in a format ready for
diagonalization. This is to allow a better 'apples to apples' comparison
to the serial SCATCI code and the fact that we are not assessing the
performance of the diagonalizers themselves. This will still test the
matrix distribution performance as the runs rely on the WriterMatrix
class which can be considered a worst case example due to the included
overhead of disk IO.

The single core build time $T_0$ is equivalent to the serial SCATCI build time.
The ideal time $T_i$ for $N_p$ processes is computed as:
\begin{equation}
\label{eq:ideal}
 T_i = \frac{T_0}{N_p}
\end{equation}
and is based on the assumption that all aspects of the calculation (including
IO) are perfectly parallel. Whilst unrealistic, it at least gives a general
sense of how the build times should scale with process count.
The update and IO time is the time taken to perform any kind of MPI
synchronization which includes the ring cycling of data in the matrix class.
Since the WriterMatrix performs a disk write in this step, this overhead is also
partially due to IO.

The node counts used were 1, 2, 3, 8 and 50. The total core counts for ARCHER are 24, 48, 72, 192 and 1200 and for Grace@UCL 16, 32, 48, 128 and 800 respectively. 
The first four tests were used to assess how the scaling behaves incrementally and the last test assesses the affect of
synchronization at overly generous core counts.

\subsection{Phosphoric acid}

Our phosphoric acid (H$_3$PO$_4$) test is based on the study of Bryjko
{\it et al} \cite{jt487}.  This is an example of an $L^2$ heavy
calculation.  The contracted portion of the matrix is only of size 712
whilst the un-contracted portion is of size 122103 giving a total
Hamiltonian size of $N=122815$. The total storage space required for the integrals was 1.5 GB. 
Using shared memory, the cost to each processor was only 64--96 MB.
The serial SCATCI reference time is $T_0=$8820~s. A serial single core run on 
MPI-SCATCI gives a time of 7800~s, a 12\% improvement, likely from the more efficient Slater rule code.

\begin{figure}
\centering
\includegraphics[width=1.22\textwidth]{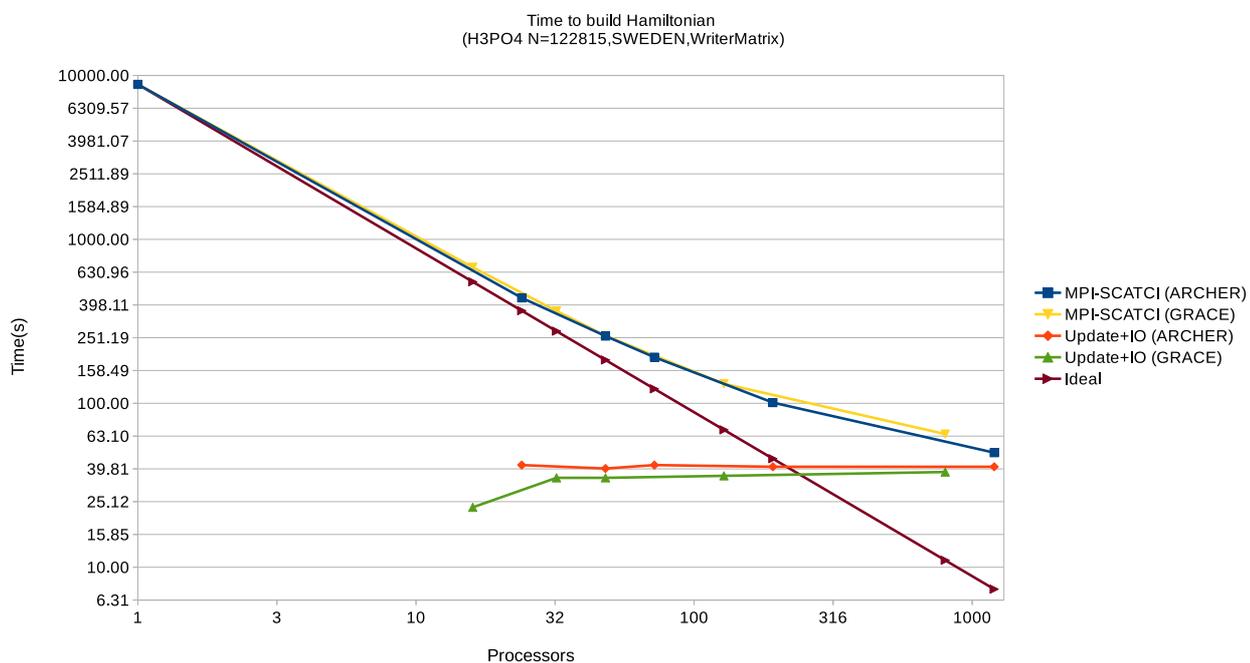}
\caption{Time taken (log scale) to build the scattering Hamiltonian for
phosphoric acid against process counts (log scale). The size of the Hamiltonian is
$N=122815$. The ideal time is computed using Eq. \ref{eq:ideal}. The
update+IO time is the time taken for MPI synchronization steps that include disk
writes by the WriterMatrix. The time taken by MPI-SCATCI includes the update+IO
time.}
\label{fig:phosphate_time}
\end{figure}

Figure \ref{fig:phosphate_time} shows the performance scaling for the
phosphoric acid calculation.
For a single node run on ARCHER, the time taken is 410~s corresponding to a
speed up of $\approx$ 21 times and a parallel efficiency of $\approx$
87\%. A Grace@UCL single node run is 630~s giving a speed up of 14 and a parallel efficiency of $\approx$
87.5\%. 
The speed up behaves linearly up to 72 cores before reducing at
192 cores and approaching the update+IO time at 1200 cores. This
reduction comes from the fact that the update time now becomes a
significant portion of the total time, reaching to $\approx$ 70\% of
total and reducing parallel efficiency to $\approx$ 11 \%. However it
is worth noting the behaviour of the update+IO time is essentially
constant as discussed previously and arises solely from the fewer
number of updates required in a single run offsetting the
communication overhead at higher node counts.

For phosphoric acid, the build phase requires between 1 to 3 nodes for maximum
efficiency and reduces the calculation from hours to minutes with higher counts
considered overkill. However a higher process count would still be beneficial if
one wishes to perform diagonalization afterwards.

\subsection{Beryllium mono-hydride}

The Beryllium mono-hydride (BeH) \cite{jt682} calculation is a contraction
heavy calculation. The contracted portion of the matrix is of dimension
10104 and the $L^2$ portion is of dimension 20563 giving a total dimension of $N
=30667$.  Whilst this matrix is significantly smaller than phosphoric
acid, it acts as a better system for assessing the scaling of the
heavier contraction calculation with its higher number of target
symmetries (4) and with $\approx 19$ target states per target
symmetry. The total storage space required for the integrals was 3.0 GB. 
Using shared memory, the cost to each processor was only 128-196 MB.
The reference serial SCATCI time is $T_0=$1993.6~s. A serial single core run on 
MPI-SCATCI gives a time of 1154~s, a 58\% improvement from SCATCI.

\begin{figure}
\centering
\includegraphics[width=1.22\textwidth]{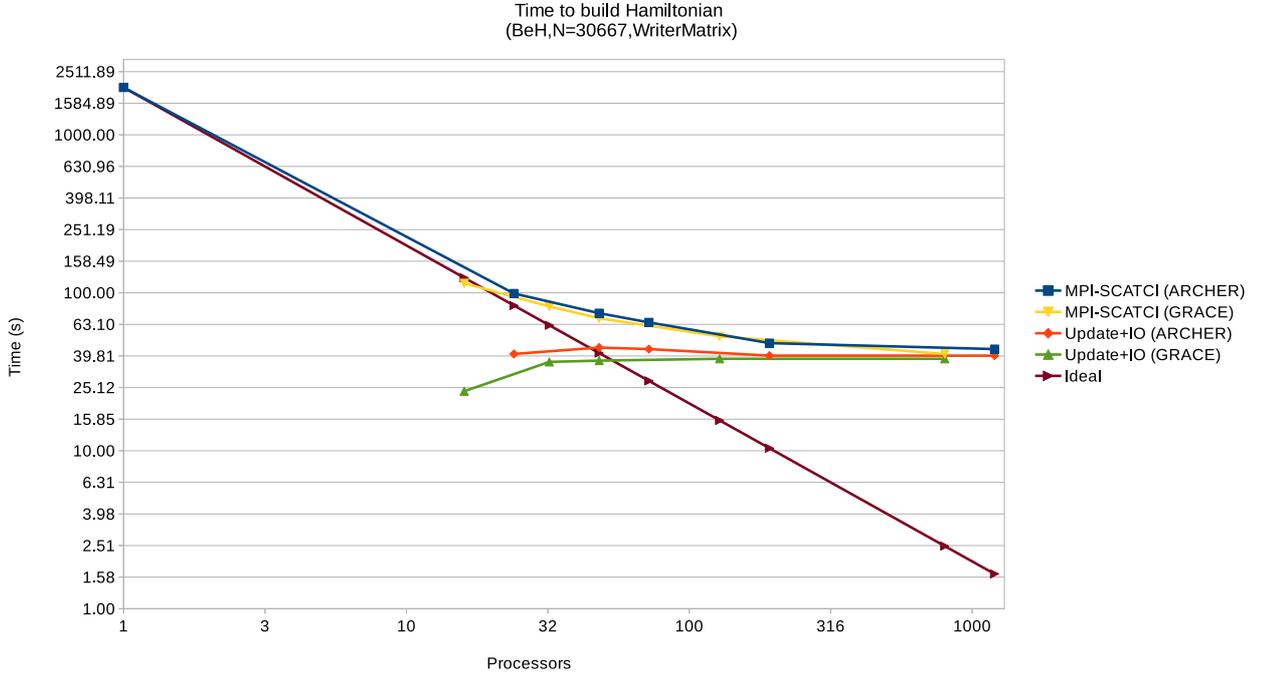}
\caption{Time taken (log scale) to build the scattering Hamiltonian for BeH
against process counts (log scale). The size of the Hamiltonian is $N =30667$.
The ideal time is computed using Eq. \ref{eq:ideal}. The update+IO time
is the time taken for MPI synchronization steps that include disk writes by the
WriterMatrix. The time taken by MPI-SCATCI includes the update+IO time.}
\label{fig:beh_time}
\end{figure}

Figure \ref{fig:beh_time} shows the performance scaling for the BeH calculation
with a reference single core time $T_0=$1993.6~s.
Almost identically to phosphoric acid, the single node ARCHER run corresponds to
a speed up of $\approx$ 20 times and a parallel efficiency of
$\approx$ 84\%. The Grace@UCL single node run gives a 17.3 times speed up giving a $>$100 \% parallel efficiency.
However, the parallel efficiency drops almost
immediately past this core count. This again arises from the update
time comprising the majority of the calculation past this point. At 1200 cores the total time has converged with the update time.
Interestingly the baseline update time also remains constant across
core counts and only increases by 10 \% from phosphoric acid. This
comes from the increased number of target symmetries that require the
prototype symbols to be synchronized for the contracted classes sans
class 4. The constant behaviour of this across high process counts may
come from the fact that the number of prototype symbols are in the
order of thousands and the parallelization reduces this to tens of
symbols for each process. These likely fit into a single packet for
the interconnect. Therefore the cost may only come from the latency of
the interconnect itself which is in the order of nanoseconds. 
Again, a single node may be considered the sweet spot for build efficiency for
BeH and higher counts benefiting diagonalization.

Grace@UCL in general has a slightly lower update time, but this is most likely due to better disk IO as its single node update time is $\approx$ 30 \% better than its $>1$ node performance.
Running a smaller scale test on BeH with 8-12 cores over 1Gbps Ethernet, update times are 91.1~s and 171~s respectively. Considering that the resulting Hamiltonian is $\approx$ 5.6 GB, the update message moving between nodes can be as big as 1 GB, oversaturating the interconnect bandwidth. Since both Infiniband interconnects are in the realm of $\approx$ 50 Gbps, and that each core is limited to a maximum of $\approx$ 2 GB of matrix elements, the Infiniband bandwidth is not fully utilized. Therefore it is most likely the IO bandwidth that is limiting the synchronization time. This is most apparent when using a SCALAPACK Matrix as it only requires a look-up and insertion into an array. For both GRACE and ARCHER, the update times for this matrix type is around $\approx$ 17.2 s.

For both examples, the overall behaviour of the code is that the
parallel efficiency is determined by the percentage of the total time
taken by the update. In a sense, the update for both matrices is
unaffected by core count.  The greatest benefit of the code may lie in
problems in the order of millions or tens of millions that take days
or months to complete. High core counts may reduce these calculations
to hours which would still remain significantly greater than the
update time.

\section{Conclusion}

The UKRMol code SCATCI has been rewritten to modern standards with MPI
integrated for large parallel build and diagonalization. It exploits
OOP paradigms to provide flexibility for future development. A
parallelization of the efficient algorithm provided by Tennyson \cite{jt180}
reduces a 3 hour-long calculation on phosphoric acid to several tens of seconds with only 1-3
nodes. It works by integrating several new parallel algorithms for
each class to exploit their particular contraction behaviour in order
to achieve a high degree of efficiency.  Additionally the code
supports LAPACK, ARPACK, SCALAPACK and SLEPc diagonalizers and has the
ability to support many more if desired with only a few lines of code.

Use of the R-Matrix with pseudo-states (RMPS) method can rapidly lead
to desirable cases where the matrix build is both large (1,000,000+)
and computationally demanding \cite{jt444}. Such calculations,
which are important to model polarization effects in a truly
{\it ab initio} manner \cite{jt468}, are the key for studying
low-lying resonances in systems such as
electron-uracil. Such calculations are currently underway.

This article has focused heavily on electron - molecule scattering
aspects of the UK Molecular R-matrix codes. In fact MPI-SCATCI can
be used to address other problems. A powerful but not greatly used
aspect of the codes is for the studies of high-lying but bound Rydberg
states of molecules. Studies have shown the use of scattering wavefunctions
provides a much more efficient means of identifying these states
than standard quantum-chemistry electronic structure calculations
\cite{jt560}. The UKRMol codes are also being increasingly used to
study photoionisation \cite{14hbm,14rhk,15bhm,jt601} and
photodetachment \cite{16kdd}. This use raises an important
technical issue with the SCATCI algorithm since the contracted
Hamiltonian is based on the use of 
very lengthy strings of effective configuration state functions (CSFs).
These CSFs, which represent entire target CI wavefunctions, do not obey
the standard Slater's rules. This means their use in computing the
transition dipole moment matrix elements required for photon-driven
processes requires special algorithms. Harvey {\it et al} \cite{14hbm}
have implemented such an algorithm for SCATCI and we anticipate MPI-SCATCI
being extensively used for future calculations on processes involving photons. 

\section{Acknowledgements}
This work was funded under the embedded CSE programme of the ARCHER UK National
Supercomputing Service (http://www.archer.ac.uk) as project eCSE08-7.
The authors acknowledge the use of the UCL Grace High Performance Computing Facility (Grace@UCL), and associated support services, in the completion of this work.
We thank Jimena Gorfinkiel and Zdenek Masin for helpful discussions, and Daniel Darby-Lewis and Kalyan
Chakrabarti for help with input files.
AFA would also like to thank Dr. Faris N. Al-Refaie, Lamya Ali, Sarfraz and Eri
Aziz, and Rory and Annie Gleeson for their support.

\bibliographystyle{elsarticle-num}

%\bibliography{journals_phys,methods,jtj,mpi-scatci,rmat,kushner}

\end{document}